\documentclass{article} 
\usepackage{iclr2021_conference,times}

\usepackage{hyperref}
\usepackage{url}

\usepackage{amssymb,amsmath,amsthm,enumerate,threeparttable,bm,subfigure,graphicx}
\usepackage{algorithm, algorithmic}
\usepackage{booktabs}
\usepackage{multirow}
\usepackage{color}
\usepackage{makecell}
\usepackage{float}
\usepackage{wrapfig}

\newcommand{\ie}{\textit{i.e.}}
\newcommand{\eg}{\textit{e.g.}}

\long\def\comment#1{}
\newtheorem{defn}{Definition}

\title{Backdoor Attack in the Physical World}

\author{Yiming Li$^{1}$, Tongqing Zhai$^{1}$, Yong Jiang$^{1}$, Zhifeng Li$^{2}$, Shu-Tao Xia$^{1}$\\
$^{1}$Tsinghua Shenzhen International Graduate School, Tsinghua University\\
$^{2}$Tencent AI Lab\\
\href{mailto:li-ym18@mails.tsinghua.edu.cn}{li-ym18@mails.tsinghua.edu.cn};
\href{mailto:xiast@sz.tsinghua.edu.cn}{xiast@sz.tsinghua.edu.cn}\\
}

\iclrfinalcopy 
\begin{document}

\maketitle

\vspace{-1em}
\begin{abstract}
Backdoor attack intends to inject hidden backdoor into the deep neural networks (DNNs), such that the prediction of infected models will be maliciously changed if the hidden backdoor is activated by the attacker-defined trigger. Currently, most existing backdoor attacks adopted the setting of \emph{static} trigger, $i.e.,$ triggers across the training and testing images follow the same appearance and are located in the same area. In this paper, we revisit this attack paradigm by analyzing trigger characteristics. We demonstrate that this attack paradigm is vulnerable when the trigger in testing images is not consistent with the one used for training. As such, those attacks are far less effective in the physical world, where the location and appearance of the trigger in the digitized image may be different from that of the one used for training. Moreover, we also discuss how to alleviate such vulnerability. We hope that this work could inspire more explorations on backdoor properties, to help the design of more advanced backdoor attack and defense methods.
\end{abstract}

\vspace{-1em}
\section{Introduction}
\vspace{-0.6em}

Recent studies showed that some regular (\ie, non-optimized) perturbations (\eg, the local patch stamped on the image) could mislead DNNs, through influencing model weights in the training process \citep{liu2020survey,li2020backdoor,gao2020backdoor}. 
It is called as {\it backdoor attack}. Specifically, some training images are modified by adding the trigger (\eg, the local patch). These modified images with the attacker-specified target label, together with 
benign training samples, are fed into the DNN model for training. Consequently, trained DNNs perform well on benign testing samples, whereas their prediction will be changed when the same trigger is contained in the attacked image. Since attacked DNNs perform normally on benign samples, it is difficult for users to realize the attack. Hence, the insidious backdoor attack is a serious threat to the practical application of DNNs.

Many backdoor attacks have been proposed through designing different types of triggers \citep{gu2017badnets,liao2018backdoor,turner2019label,zhao2020clean,li2020backdoor1,zhai2021backdoor}. Currently, most existing works adopted the setting of \emph{static} trigger, where the triggers across the training and testing images are the same. However, the location and appearance of the trigger in the digitized image may be different from that of the one used for training in the physical world. It raises an intriguing question: {\it When the trigger in the attacked testing image is different from that used in training, can it still activate the hidden backdoor?}

To answer this question, we explore the impacts of two basic characteristics of the trigger, including \emph{location} and \emph{appearance}. We demonstrate that if the location or appearance is slightly changed, then the attack performance may degrade sharply. It reveals that attacks with the static trigger pattern may be non-robust to the change of trigger. The above observation inspires two further questions:

{\bf (1)} {\it Can we utilize this non-robustness to defend existing backdoor attacks?} {\bf (2)} {\it How to enhance the performance of existing backdoor attacks, such that they are robust to the change of trigger?}

In this work, we propose a simple yet effective defense towards attacks with the static trigger pattern in which the testing sample is transformed (\eg, flipping or scaling) before the prediction. The transformation is a feasible approach to change the trigger's location and appearance. Besides, we propose to enhance the transformation robustness of attacks that all poisoned images will be randomly transformed before feeding into the training process. This enhancement could be naturally combined with any backdoor attack. Moreover, we demonstrate the connection between the proposed attack enhancement and the physical attack, which implies that enhanced attacks could still succeed in the physical world whereas standard backdoor attacks will fail. 

\comment{
The main contributions of this work are three-fold: \textbf{(1)} We demonstrate that the location and appearance of triggers have crucial impacts on activating the backdoor. \textbf{(2)} We verify that attacks with the static trigger pattern are transformation vulnerable, which inspires a simple yet effective defense method. \textbf{(3)} We propose an effective method to enhance the robustness of existing attacks against the change of trigger, and connect the proposed enhancement with the physical attack. 
}

\begin{figure}[ht]
\vspace{-2em}
\begin{minipage}[b]{0.5\linewidth}
    \centering
    \subfigure[VGG-19]{
    \includegraphics[width=0.48\textwidth]{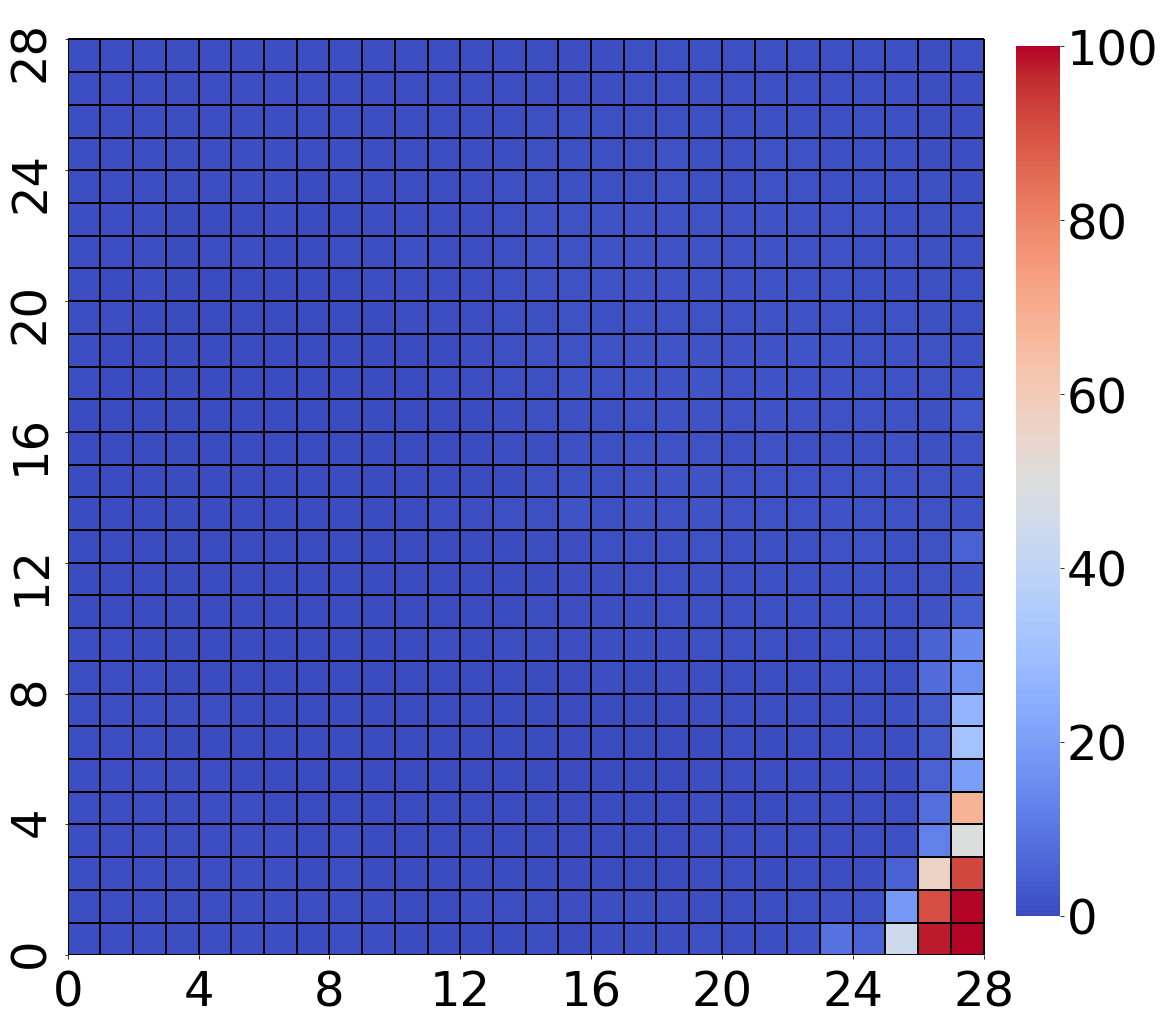}}
    \subfigure[ResNet-34]{
    \includegraphics[width=0.48\textwidth]{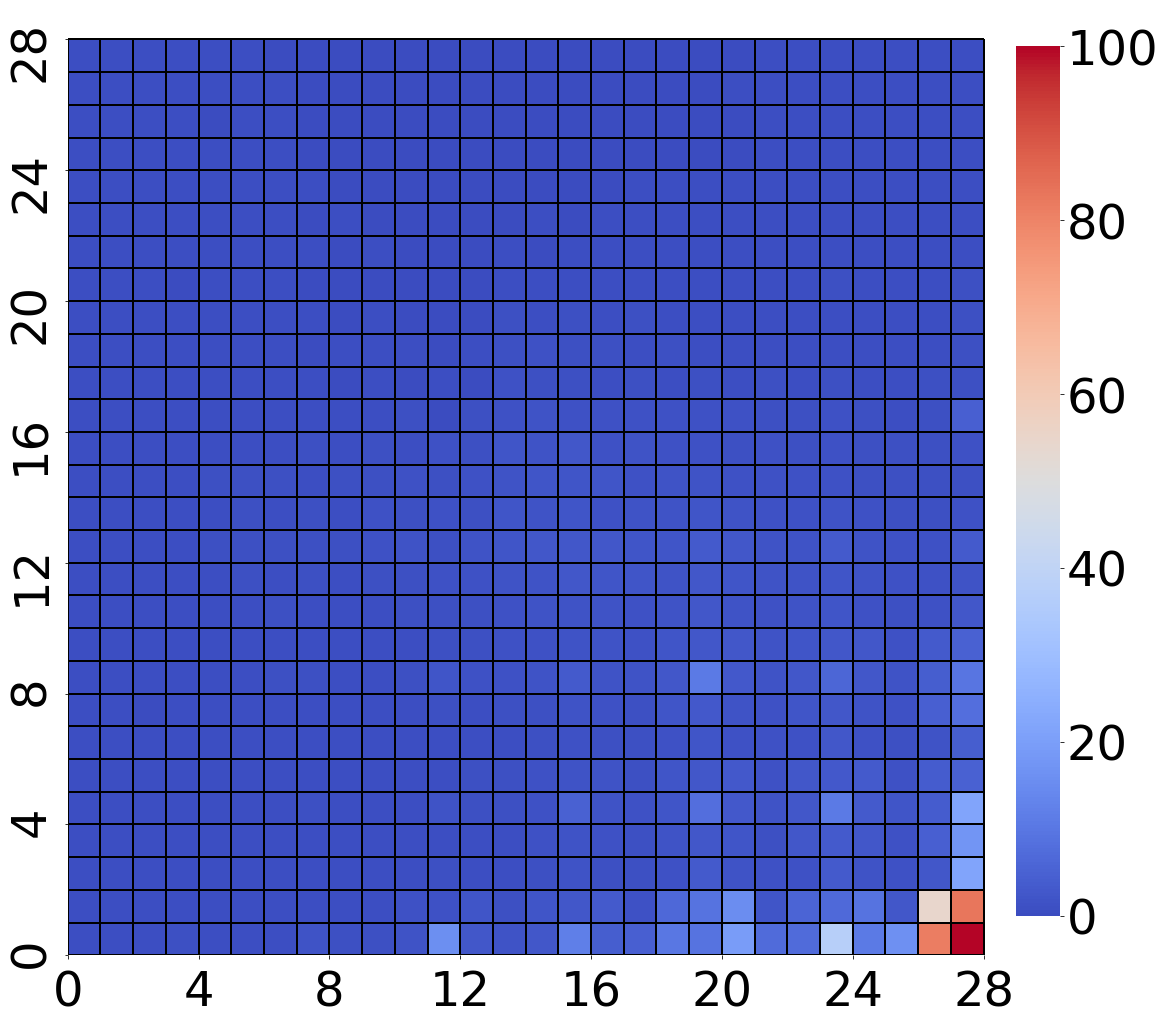}}
    \vspace{-0.6em}
    \caption{The heatmap of the attack success rate when the trigger is in different position at attacked images. The right corner is the position of the trigger in the poisoned images used for training.}
\label{fig_position}
\end{minipage}\quad
\begin{minipage}[b]{0.47\linewidth}
    \centering
    \includegraphics[width=0.9\textwidth]{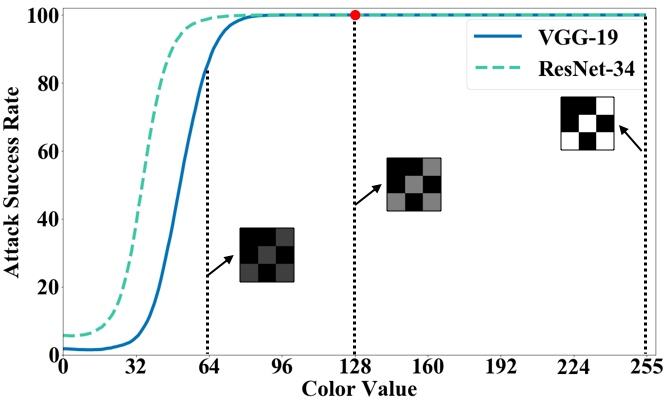}
    \vspace{-0.6em}
    \caption{ASR and appearance of the trigger with different non-zero color value in attacked images. The red dot indicates the ASR of trigger with original color value (128 pixels).}
    \label{fig_color}
\end{minipage}
\vspace{-1.2em}
\end{figure}

\vspace{-0.8em}
\section{The Property of Existing Attacks with Static Trigger}
\vspace{-0.6em}
\label{sec_limit}

\subsection{Backdoor Attack with Static Trigger}
\vspace{-0.4em}
We consider the scenario that the user cannot fully control the training process of the model $C(\cdot;w)$. Let $y_{target}$ denotes the target label, $\mathcal{D}_{train} = \{ (\bm{x}, y) \}$ indicates the (benign) training set. The target of backdoor attack is to obtain an \emph{infected model}, which performs well on benign tesing images whereas it may have been injected some insidious backdoors.

Generating poisoned images is the first step of backdoor attacks. Specifically, the poisoned image $\bm{x}_{poisoned}$ is generated through a generation $G$ based on the trigger $\bm{x}_{trigger}$ and the benign image $\bm{x}$, $e.g.$,
$
    \bm{x}_{poisoned} = G(\bm{x};\bm{x}_{trigger}) = (\bm{1}-\bm{\alpha}) \otimes \bm{x} + \bm{\alpha} \otimes \bm{x}_{trigger},  
    \label{eq: poisoned image}
$
where $\bm{\alpha} \in [0,1]^{C \times W \times H}$ is a trade-off hyper-parameter and $\otimes$ indicates the element-wise product. After that, all generated poisoned samples $\mathcal{D}_{poisoned}=\{(\bm{x}_{poisoned}, y_{target})\}$ and a set of benign samples $\mathcal{D}_{benign}$ will be used for training the model $C(\cdot;w)$, $i.e.$, 
$
    \min_{w} \mathbb{E}_{(x,y) \in \mathcal{D}_{poisoned} \cup \mathcal{D}_{benign}}  \mathcal{L}\left(C(\bm{x};w), y\right),
$
where $\mathcal{L}(\cdot)$ indicates the loss function, such as the cross entropy.

\vspace{-0.5em}
\subsection{The Effects of Different Characteristics}
\begin{figure}[ht]
\begin{minipage}[b]{0.55\linewidth}
\begin{defn}[Minimum Covering Box]
The minimum covering box is defined as the minimum bounding box in the poisoned image covering the whole trigger pattern ($i.e.$, all non-zero $\bm{\alpha}$ entries).
\end{defn}

\vspace{-0.2em}

\begin{defn}[Two Characteristics of Backdoor Trigger]
\label{def: three characteristics}

A trigger can be defined by two independent characteristics, including \textbf{location} and \textbf{appearance}. Specifically, \textbf{location} is defined by the position of the pixel at the bottom right corner of the minimum covering box, and \textbf{appearance} is indicated by the color value and the specific arrangement of pixels corresponding to non-zero $\bm{\alpha}$ entries in the minimum covering box.
\end{defn}
\end{minipage}
\hfill
\begin{minipage}[b]{0.41\linewidth}
 \centering
 \vspace{-0.5em}
 \includegraphics[width=\textwidth]{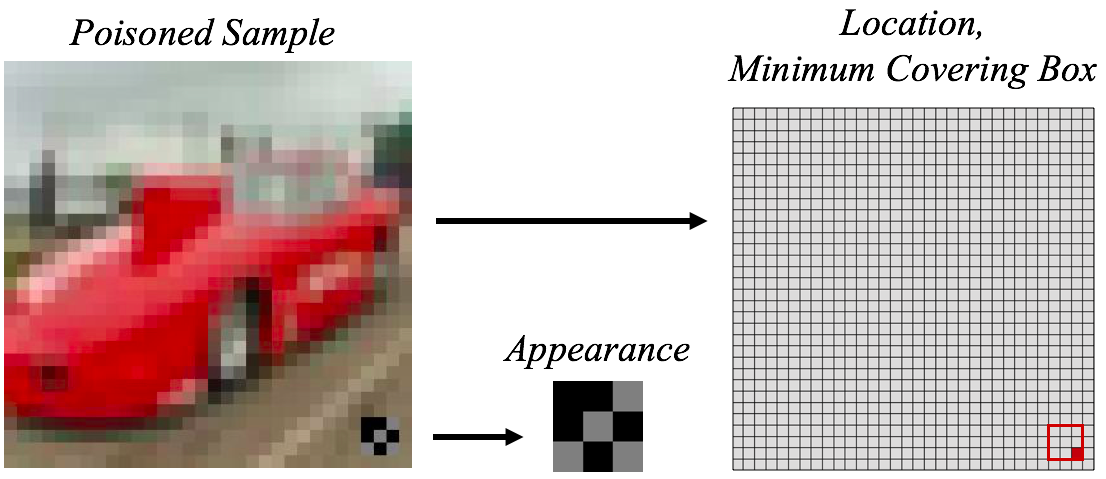}
 \vspace{-1.2em}
 \caption{The illustration of characteristics of the backdoor trigger. The red box represents the boundary of the minimum covering box, and the red pixel indicates the trigger location.} 
 \label{ill_TwoChars}
\end{minipage}
\vspace{-0.8em}
\end{figure}

\vspace{-0.5em}
\label{sec_char}
One backdoor trigger can be specified by two independent characteristics, including \emph{location} and \emph{appearance}, as defined in Definition \ref{def: three characteristics}. In this section, we study their individual effects.

\vspace{-0.4em}

\noindent \textbf{Settings.} We adopt BadNets \citep{gu2019badnets} as an example to study their effects. Specifically, we use VGG-19 \citep{simonyan2014very} and ResNet-34 \citep{he2016deep} as the model structure, and conduct experiments on CIFAR-10 dataset \citep{krizhevsky2009learning}. The trigger is a $3\times3$ black-gray square, as shown in Figure \ref{ill_TwoChars}. We adopt the \emph{attack success rate} (ASR), which is defined as the accuracy of attacked images predicted by the infected classifier, to evaluate the attack performance.

\noindent \textbf{The Effect of Location.}
While preserving the appearance of the trigger, we change its location in inference process to study its effect to the attack performance. As shown in Figure \ref{fig_position}, when moving the location with a small distance ($e.g.$, $2 \sim 3$ pixels), the ASR will drop sharply from $100\%$ to below $50\%$. It tells that the attack performance is sensitive to the location of the backdoor trigger.

\noindent \textbf{The Effect of Appearance.} 
While keeping the location of the trigger, we change its appearance in the inference stage to study its effect to the attack performance. The trigger appearance could be modified by changing the shape or the pixel values. For the sake of simplicity, here we only consider the change of pixel values. 
Specifically, there are only two values of the pixels within the trigger, $i.e.$, 0 and 128. 
We change the value 128 to different values from 0 to 255. As shown in Figure \ref{fig_color}, the ASR degrades sharply along with the decreasing of non-zero pixel values, while is not significantly influenced when the values are increased. 
According to this simple experiment, it is difficult to describe the exact relationship between the appearance and the attack performance since the change modes of appearance are rather diverse. However, it at least tells that the attack is sensitive to the trigger appearance. More explorations about this phenomenon will be discussed in our future work.

\vspace{-0.7em}
\section{Transformation-inspired Defense and Attack Enhancement}
\vspace{-0.6em}

\vspace{-0.1em}
\subsection{Backdoor Defense via Transformations}
\vspace{-0.4em}
\label{sec_transdefense}
Since the user doesn't have the information about the trigger, it is impossible to exactly manipulate it in the inference process. Instead, we propose a transformation-based defense by changing the whole image with some transformations (\eg, flipping or scaling), as shown in Definition \ref{def_defense}.

\begin{defn}[Transformation-based Defense]\label{def_defense}
    The transformation-based defense is defined as introducing a transformation-based pre-processing module on the testing image before prediction, $i.e.$, instead of predicting $\bm{x}$, it predicts $T(\bm{x})$, where $T(\cdot)$ is a transformation. 
\end{defn}
\vspace{-0.6em}

This simple strategy enjoys several advantages: {\bf (1)} it is efficient since it only requires to transform the testing image; {\bf (2)} it is attack-agnostic, therefore it can defend different attacks simultaneously; {\bf (3)} it is data-free and model-free, $i.e.,$ the defender does not need to have any additional samples or modify the model. Therefore it would be the primary choice when adopting third-party model APIs.

\vspace{-0.5em}
\subsection{Transformation-based Enhancement and Physical Backdoor Attack}
\vspace{-0.4em}
\label{sec_enhancement}

Once transformations adopted by the user/defender are known, it would be easy to design an adaptive attack by introducing those transformations in the training process. However, attackers usually have no information about the inference process. To tackle this difficulty, we propose to approximate them with a set of widely adoped transformations ${T_i(\cdot;\theta_i)}$. For each $T_i$, we define a value domain $\Theta_i$ for $\theta_i$. $\Theta_i$ is parameterized by the maximal transformation size $\epsilon_i$, 
$i.e.$,
$
    \Theta_i = \{\theta|dist_i(\theta, I) \leq \epsilon_i\}, 
$
where $dist_i(\cdot, \cdot)$ is a given distance metric for $T_i$ and $I$ indicates the identity transformation.

Consequently, the (compound) transformation used in the enhanced attack is specified as $\mathcal{T}= \{T(\cdot;\bm{\theta})|\bm{\theta} \in \prod_{i=1}^{n}\Theta_i\}$. 
Then, the training objective of the enhanced attack is formulated as 
\begin{equation}\label{obj_attack_class_enhance}
    \min_{w} \mathbb{E}_{\bm{\theta}} \left[\mathbb{E}_{(\bm{x},y) \in \mathcal{D}_{poisoned}^{(T(\cdot;\bm{\theta}))} \cup \mathcal{D}_{benign}} \left[\mathcal{L}\left(C(\bm{x};w), y\right)\right]\right].
\end{equation}

To solve the problem (\ref{obj_attack_class_enhance}) exactly, attackers need to conduct the training process with all possible transformed variants, which is computation-consuming. Instead, 
we propose a sampling-based method where we sample only one configuration, \ie, $\bm{\theta} \sim \prod_{i=1}^{n}\Theta_i$ to transform each poisoned image in each time. Then, we use the transformed poisoned images and benign images for training.

\textbf{Connecting the proposed attack enhancement and physical attack. } In real-world scenarios, the testing image may be acquired by some digitizing devices. As such, the trigger in the digitized image may be different from the one used for training. These differences can be approximated by some widely used transformations (\eg, spatial transformations), which have been incorporated into the proposed attack enhancement. Thus, it is expected that attacks with the proposed enhancement can still be effective in the physical world, which will be futher verified in Section \ref{sec:attack_enhance}.

\vspace{-1.1em}
\section{Experiment}
\vspace{-0.8em}
\label{sec_exp}

\subsection{Transformation-based Defense}
\label{transdefense}
\vspace{-0.6em}

\textbf{Settings. } We use three representative backdoor attacks, including BadNets \citep{gu2017badnets}, Blended Attack \citep{chen2017targeted}, and Consistent Attack \citep{turner2019label} to evaluate the performance of backdoor defenses. 
We examine two simple spatial transformations, including left-right flipping (dubbed {\it Flip}), and padding after shrinking (dubbed {\it ShrinkPad}). Specifically, ShrinkPad consists of shrinking (based on bilinear interpolation) with a few pixels ($i.e.$, shrinking size), and random zero-padding around the shrunk image. For defense comparison, we select four important baseline, including fine-pruning \citep{liu2018fine}, neural cleanse \citep{wangneural}, auto-encoder based defense (dubbed Auto-Encoder) \citep{liu2017neural}, and standard training (dubbed Standard).

\begin{table*}[ht]
\center
\scriptsize
\vspace{-2em}
\caption{Comparison of different backdoor defenses on CIFAR-10 dataset. `Clean' and `ASR' indicates the accuracy (\%) and attack success rate (\%) on testing set, respectively. The boldface indicates the best results among all preprocessing based defenses.}
\label{defense}
\scalebox{0.91}{
\begin{tabular}{c|cc|cc|cc|cccccc}
\hline
\multicolumn{1}{c|}{Model Architectures $\rightarrow$} & \multicolumn{6}{c|}{VGG-19}                                                                                                                                                                       & \multicolumn{6}{c}{ResNet-34}                                                                                                                                                                                                              \\ \hline
\multicolumn{1}{c|}{Attack Methods $\rightarrow$} & \multicolumn{2}{c|}{BadNets}                                    & \multicolumn{2}{c|}{Blended Attack}                            & \multicolumn{2}{c|}{Consistent Attack}                         & \multicolumn{2}{c|}{BadNets}                                                        & \multicolumn{2}{c|}{Blended Attack}                                                 & \multicolumn{2}{c}{Consistent Attack}                          \\ \cline{2-13} 
\multicolumn{1}{c|}{Defense Methods $\downarrow$} & Clean & \begin{tabular}[c]{@{}c@{}}ASR\end{tabular} & Clean & \begin{tabular}[c]{@{}c@{}}ASR\end{tabular} & Clean & \begin{tabular}[c]{@{}c@{}}ASR\end{tabular} & Clean & \multicolumn{1}{c|}{\begin{tabular}[c]{@{}c@{}}ASR\end{tabular}} & Clean & \multicolumn{1}{c|}{\begin{tabular}[c]{@{}c@{}} ASR \end{tabular}} & Clean & \begin{tabular}[c]{@{}c@{}} ASR \end{tabular} \\ \hline
Standard          & 91.9  & 100                                                     & 91.5     & 100                                                   & 91.3  & 95.6                                                   & 94.1  & \multicolumn{1}{c|}{100}                                                   & 93.1     & \multicolumn{1}{c|}{100}                                                   & 93.1  & 98.7                                                   \\ \hline

Fine-Pruning          & 91.3  & 0.7                                                     & 83.6     & 0.2                                                   & 72.6  & 0.1                                                   & 92.1  & \multicolumn{1}{c|}{0}                                                   & 91.9     & \multicolumn{1}{c|}{0.3}                                                   & 92.0  & 18.9                                                   \\
Neural Cleanse        & 83.3     & 0.6                                                       & 90.6     & 0.4                                                      & 86.4     & 0.7                                                      & 91.4     & \multicolumn{1}{c|}{0.7}                                                      & 91.4     & \multicolumn{1}{c|}{0.5}                                                      & 91.2     & 1.4                                                      \\ \hline
Auto-Encoder           & 86.4  & 2.1                                                     & 86.0     & 1.7                                                      & 85.4  & \textbf{2.3}                                                    & 87.5  & \multicolumn{1}{c|}{2.7}                                                    & 87.2     & \multicolumn{1}{c|}{1.9}                                                      & 88.4  & \textbf{2.1}                                                    \\ 
Flip (Ours)                 & \textbf{91.0}    & \textbf{1.1}                                                     & \textbf{91.1}     & \textbf{0.9}                                                      & \textbf{90.5}  & 95.7                                                   & \textbf{93.6}  & \multicolumn{1}{c|}{\textbf{0.8}}                                                    & \textbf{92.8}     & \multicolumn{1}{c|}{\textbf{0.8}}                                                      & \textbf{92.3}  & 98.8                                                   \\
ShrinkPad-4 (Ours)         & 87.6  & 1.6                                                    & 88.3     & 1.8                                                      & 87.5  & 3.7                                                    & 91.4  & \multicolumn{1}{c|}{1.5}                                                   & 90.6     & \multicolumn{1}{c|}{1.8}                                                      & 89.9  & 4.8                                                   \\ \hline
\end{tabular}
}
\vspace{-2em}
\end{table*}

\begin{table*}[ht]
\center
\scriptsize
\caption{The comparison between standard backdoor attacks and enhanced backdoor attacks from the aspect of attack success rate against different transformation-based defenses. }
\begin{tabular}{c|cccc|cccc}
\hline
Model Architectures $\rightarrow$                 & \multicolumn{4}{c|}{VGG-19}                                                       & \multicolumn{4}{c}{ResNet-34}                                                    \\ \hline
                        Attacks $\downarrow$, Defenses $\rightarrow$   & Standard     & Flip           & ShrinkPad-2  & ShrinkPad-4         & Standard     & Flip           & ShrinkPad-2  & ShrinkPad-4  \\ \hline
BadNets                       & \textbf{100.0} & 1.1            & 22.7           & 1.6                      & \textbf{100.0} & 0.8            & 14.9            & 1.5           \\
BadNets+             & \textbf{100.0} & \textbf{100.0} & \textbf{100.0} & \textbf{100.0}  & \textbf{100.0} & \textbf{100.0} & \textbf{100.0} & \textbf{100.0} \\ \hline
Blended Attack               & \textbf{100.0} & 0.9            & 40.8            & 1.8                      & \textbf{100.0} & 0.8            & 18.2            & 1.8            \\
Blended Attack+     & 99.9           & \textbf{99.9}  & \textbf{100.0} & \textbf{98.7}   & \textbf{100.0} & \textbf{100.0} & \textbf{100.0} & \textbf{99.5}  \\ \hline
Consistent Attack            & \textbf{95.6}           & \textbf{95.7}           & 67.1           & 3.7                      & \textbf{98.7}           & \textbf{98.8}          & 24.2           & 4.8            \\
Consistent Attack+    & 86.0     & 86.3     & \textbf{97.2}     & \textbf{90.9}     & 96.4     & 97.3     & \textbf{97.4}     & \textbf{98.7}     \\ \hline
\end{tabular}
\label{tab_EnhancedAttack}
\vspace{-2em}
\end{table*}

\textbf{Results. } As shown in Table \ref{defense}, our method is effective. Specifically, ShrinkPad with 4 pixels shrinking size could decrease the ASR by more than $90\%$ in all cases. Flip also shows satisfied defense performance towards BadNets and Blended attacks. But it doesn't work on defending against Consistent Attack since its trigger is symmetric. Compared with the state-of-the-art preprocessing based method ($i.e.,$ Auto-Encoder), the proposed method has higher clean accuracy and lower ASR in general. Besides, its performance is even on par with Fine-Pruning and Neural Cleanse, which require stronger defensive capabilities (\ie, modify the model parameters and access to benign samples).

\vspace{-0.4em}
\subsection{Attack Enhancement} \label{sec:attack_enhance}
\vspace{-0.3em}
\noindent \textbf{Resistance to Transformation-based Defense. } 
In the enhanced backdoor attack, we adopt random Flip followed by random ShrinkPad in the random transformation layer. There is only one hyper-parameter in the enhanced attack, $i.e.$, the maximal shrinking size, which is set to 4 pixels. Other settings are the same as those used in Section \ref{transdefense}. As demonstrated in Table \ref{tab_EnhancedAttack}, enhanced backdoor attacks can still achieve a high ASR even under the defenses with spatial transformations. Specifically, the ASR of enhanced backdoor attacks is better than the one of their corresponding standard attack under defenses in almost all cases. The only exception is the Consistent Attack+ under Flip defense. It is partially due to the fact the trigger of Consistent Attack is symmetrical, as mentioned in Section \ref{transdefense}. Besides, compared to BadNets+ and Blended Attack+, Consistent Attack+ poisoned fewer images (see the attack settings), which is not favorable to the random trigger.

\comment{
\vspace{-0.4em}
\begin{figure}[ht]
\begin{minipage}[b]{0.43\linewidth}
\noindent \textbf{Backdoor Attack in the Physical World. } In this section, we verify the effectiveness of the proposed enhancement in the physical world. Since trigger stamping instead of pixel-wise manipulation would more possibly happen in real-world applications, we compare BadNets and BadNets+ in this experiment. We randomly pick some testing images on the CIFAR-10 dataset with the trigger to take pictures with differently relative location (near and far), as shown in Figure \ref{phy1}. In the results of all figures, BadNets+ successfully enforces the prediction to the target label, while BadNets fails. 
\end{minipage}
\hfill
\begin{minipage}[b]{0.55\linewidth}
 \centering
 \includegraphics[width=\textwidth]{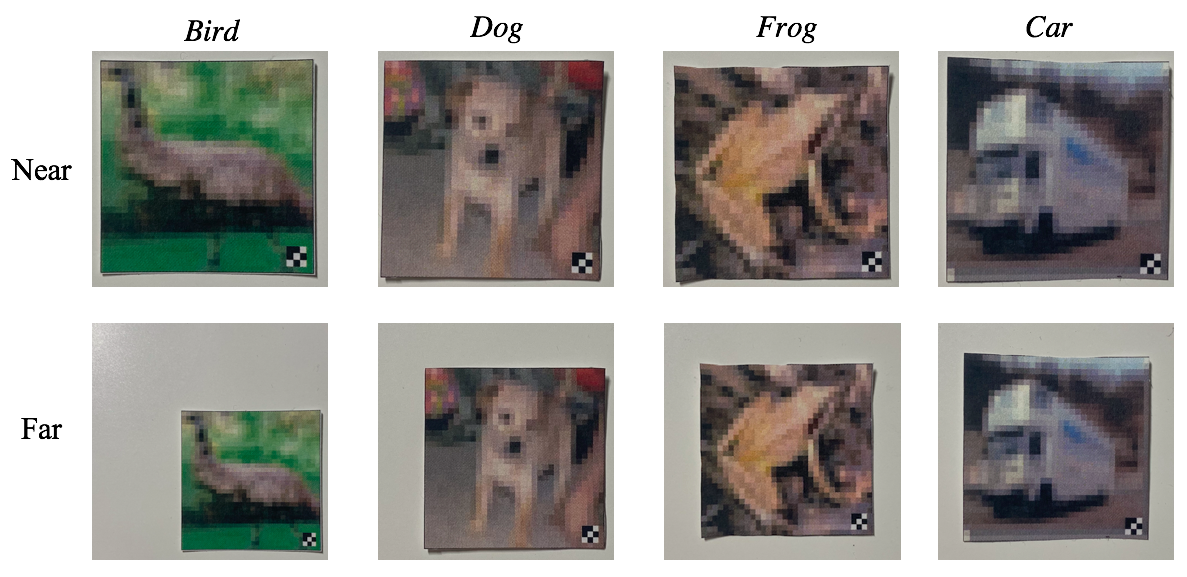}
 \vspace{-1.5em}
 \caption{Some printed CIFAR-10 images taken by a camera with different distances. }
 \label{phy1}
\end{minipage}
\vspace{-0.8em}
\end{figure}
}

\begin{wrapfigure}{r}{8cm}
\centering
\includegraphics[width=0.45\textwidth]{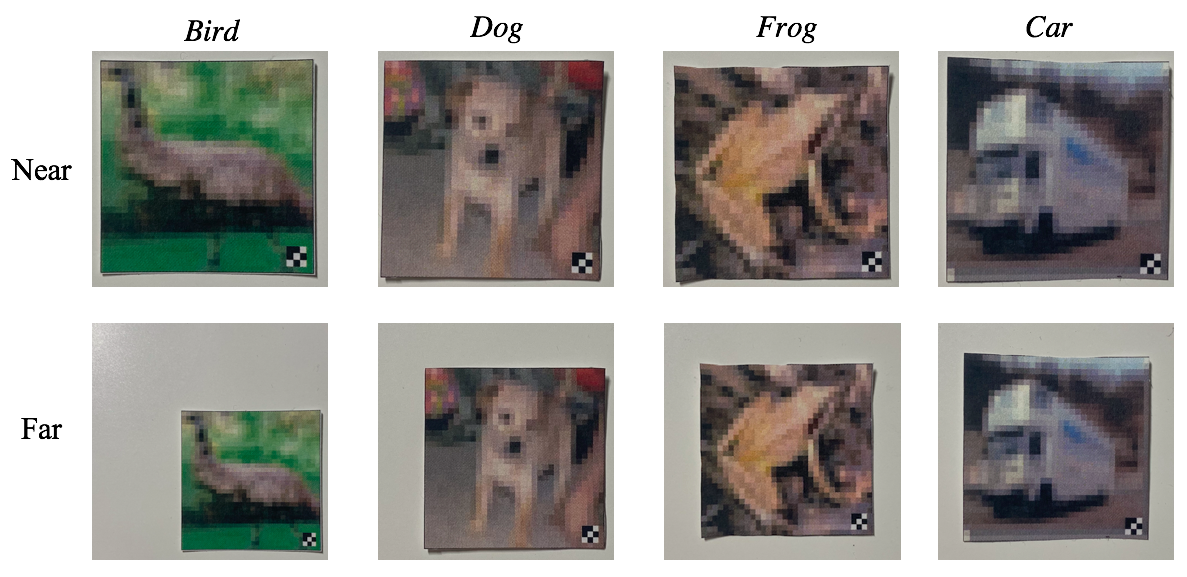}
\vspace{-1.2em}
\caption{Some printed CIFAR-10 images taken by a camera with different distances. }
\label{phy1}
\end{wrapfigure}

\noindent \textbf{Attack in the Physical World. } In this section, we verify the effectiveness of our attack enhancement in the physical world. Since patch stamping instead of pixel-wise manipulation would more possibly happen in real-world applications, we compare BadNets and BadNets+ in this experiment. We randomly pick some attacked samples on CIFAR-10 to take pictures with differently relative location (near and far), as shown in Figure \ref{phy1}. BadNets+ successfully enforces the prediction of all figures to the target label, whereas BadNets fails. These results verify the connection between our enhancement and the physical attack, as stated in Section \ref{sec_enhancement}.

\vspace{-0.5em}
\section{Conclusion}
\vspace{-0.4em}
In this paper, we explore the property of backdoor attacks. We reveal that existing attacks are mostly transformation vulnerable. We propose a transformation-based enhancement to reduce the vulnerability and link the proposed enhancement to the physical attack. We hope that our approach could inspire more explorations on backdoor properties, to help the design of more advanced methods.


\bibliography{iclr2021_conference}
\bibliographystyle{iclr2021_conference}

\end{document}